\documentclass{article}
\usepackage{cite}
\usepackage{inputenc}
\usepackage[english]{babel}
\usepackage[nohead,includefoot,margin=2cm]{geometry}
\usepackage[compact,raggedright,small]{titlesec}
\usepackage[affil-it]{authblk}
\usepackage[runin]{abstract}
\usepackage{multicol}
\usepackage{graphicx}
\usepackage{float}
\usepackage{subfloat}
\usepackage{subfig}

\newcommand{\Ni}{Ni(II) acetylacetonate}

\addto{\Affilfont}{\small}

\title{\large\bfseries Near-field optical investigation of Ni clusters inside single-walled carbon nanotubes on the nanometer scale}
\date{\small (Dated \today)}
\author[1]{Gergely N\'emeth}
\author[1]{D\'aniel Datz}
\author[1]{\'Aron Pekker}
\author[2]{Takeshi Saito}
\author[3]{Oleg Domanov}
\author[3,4]{Hidetsugu Shiozawa}
\author[5]{S\'andor Lenk}
\author[6]{B\'ela P\'ecz}
\author[5]{P\'al Koppa}
\author[1]{Katalin Kamar\'as}
\affil[1]{Institute for Solid State Physics and Optics, Wigner Research Centre for Physics, Hungarian Academy of Sciences, Konkoly Thege u. 29-33, H-1121 Budapest, Hungary}
\affil[2]{Nanomaterials Research Institute, AIST, 1-1-1 Higashi, Tsukuba 305-8565, Japan}
\affil[3]{Faculty of Physics, University of Vienna, Boltzmanngasse 5, A-1090 Vienna, Austria}
\affil[4]{J. Heyrovsky Institute of Physical Chemistry, Academy of Sciences of the Czech Republic, Dolejskova 3, CZ-182 23 Prague 8, Czech Republic}
\affil[5]{Department of Atomic Physics, Budapest University of Technology and Economics, Budafoki \'ut 8, H-1111 Budapest, Hungary}
\affil[6]{Institute for Technical Physics and Materials Science, Centre for Energy Research, Hungarian Academy of Sciences, Konkoly Thege u. 29-33, H-1121 Budapest, Hungary}

\begin{document}
   \maketitle

   \begin{abstract}
     We used scattering-type scanning near-field optical microscopy (s-SNOM) to characterize nickel nanoclusters grown inside single-walled carbon nanotubes (SWCNT). The nanotubes were filled with Ni(II) acetylacetonate and the molecules were transformed into nickel clusters via annealing. The metal clusters give high local contrast enhancement in near-field phase maps caused by the excitation of free charge carriers. The near-field contrast was simulated using the finite dipole model, approximating the clusters with elliptical nanoparticles. Compared to magnetic force microscopy, s-SNOM appears much more sensitive to localize metal clusters inside carbon nanotubes. We estimate the detection threshold to be $\sim$ 600 Ni atoms.
   \end{abstract}
	

\section{Introduction}
One of the unique applications of carbon nanotubes is their use as nanocontainers for various encapsulated species. Nanoscale metal clusters present a special perspective in this regard as the tubes give both a natural constraint and an effective protection from the environment. Following the early description of the filling procedure\cite {sloan98}, a considerable variety of both single- and multiwalled nanotubes combined with several metals were produced (for a review, see Ref. \cite{soldano15}), and their structural \cite{guan08}, magnetic \cite{lutz10} and superconducting \cite{jankovic06,tombros08} properties investigated. Metallocenes in single-walled carbon nanotubes (SWCNTs) \cite{li05b} represent a special class among these hybrid systems, as they constitute nanoreactors for both metal cluster formation \cite{li06} and inner nanotube growth \cite{shiozawa08}. In a  recent study \cite{shiozawa15}, upon annealing nickelocene encapsulated in SWCNTs, superparamagnetic nickel clusters were formed that are considered as high performance single domain magnets with high coercivity. Here we measure such clusters by scattering-type near-field optical microscopy (s-SNOM) to both probe the metallicity of such small nanostructures and to establish the sensitivity of the method.

\section{Sample preparation and characterization}
Details of encapsulation of \Ni\ in e-Dips single walled carbon nanotubes with tube diameters of $1.7\pm 0.1nm$ are given in Ref. \cite{shiozawa15}. The encapsulated molecules were transformed to nickel clusters by annealing. The size of these clusters can be controlled by the annealing temperature: In order to get fewer but well separated, long clusters we heated the sample in vacuum at $700^{\circ}C$ for 2 hours. Previous results  using similar conditions \cite{shiozawa15} showed the formation of nickel clusters with aspect ratio ranging approximately from 1 to 15. Near-field microscopy was performed on samples deposited on a silicon substrate by vacuum filtration \cite{wu04}.

In order to follow and control the annealing process we verified the disappearance of \Ni\ molecules using attenuated total reflection (ATR) spectroscopy in the mid-infrared (MIR) region (Fig. S1). The disappearance of the acetylacetonate-related peaks indicates the successful decomposition of the molecules and the possible formation of Ni clusters.

We also observed nickel clusters being created inside the nanotubes via transmission electron microscopy (TEM). The nanotubes were dispersed in toluene, sonicated for 1 hour and collected onto a TEM grid with ultrathin carbon film, then images were taken using both a JEOL 3010 and an FEI THEMIS microscope. A typical image is shown in Fig. S2. The clusters look spatially separated enough to enable the measurement of the optical response via s-SNOM where the possible resolution limit is around 20 nm.

\section{Near-field measurements}
\begin{figure}[h!]
\begin{center}
\includegraphics[width=0.5\linewidth]{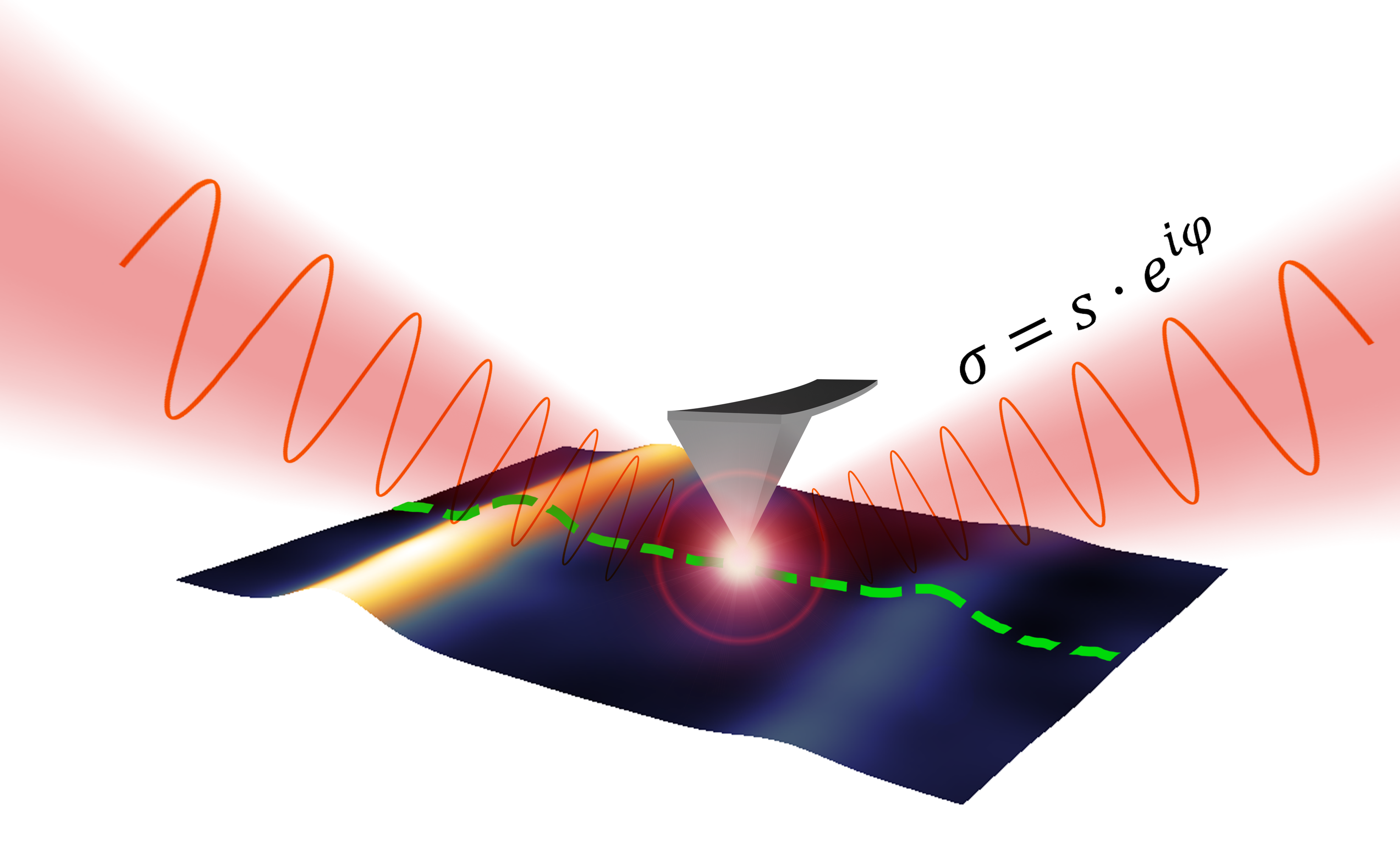}
\caption{Illustration of the near-field scattering process and the illuminated probing tip.}
\label{fig:nanofocus}
\end{center}
\end{figure}

We applied scattering-type near-field optical microscopy (s-SNOM) to image nano-sized nickel atom clusters inside single walled carbon nanotubes based on their infrared optical response with spatial resolution beyond the diffraction limit. The s-SNOM setup utilizes an atomic force microscope with a metal-coated tip illuminated from the side by a focused laser beam. (In our case, the source was an infrared (980 cm$^{-1}$) quantum cascade laser (QCL)). The illuminated metal tip acts as an optical antenna \cite{Bharadwaj2009} and enhances the electric field under the tip as depicted in Fig. \ref{fig:nanofocus}. The extension of this well-localized, high amplitude electric field depends on the tip apex radius \cite{Hillenbrand2002}. As this nano-sized light probe is scanned in the proximity to the surface, an optical interaction occurs between the probe and the sample.

This interaction results in propagating waves via scattering from the volume below the probing tip. The amplitude and the phase of the scattered wave are in close relation with the local optical properties of the sample \cite{Hillenbrand2000,Stiegler2011}. The very weak near-field scattered light is then demodulated at the higher harmonics of the tip oscillation frequency and further analyzed using pseudo-heterodyne detection \cite{Hillenbrand2006} based on a Michelson-type interferometer, shown in Fig. S3.

This complex setup enables the simultaneous measurement of the sample topography and both the amplitude and the phase of the near-field scattered light. The s-SNOM microscopy yields very high wavelength-independent spatial resolution $(\approx 20nm)$ and high optical response that gives the opportunity to study nanostructures consisting of only a few hundred atoms.

Magnetic force microscopy (MFM) was also performed with a separate AFM instrument (Bruker Dimension Icon) using a standard Bruker MESP magnetic AFM tip.

\section{Theoretical predictions}
In the applied mid-infrared spectral region the near-field signal originates from the excitation of free charge carriers. This effect is ideal for characterizing metallic components as they produce high near-field signal contrast compared to the silicon substrate. Although carbon nanotubes show near-field contrast themselves \cite{nemeth16,nemeth17}, their conductivity is negligible compared to real metals like nickel, therefore we do not expect observable contribution from the nanotube walls.

\begin{figure}[h!]
\begin{center}
\includegraphics[width=0.4\linewidth]{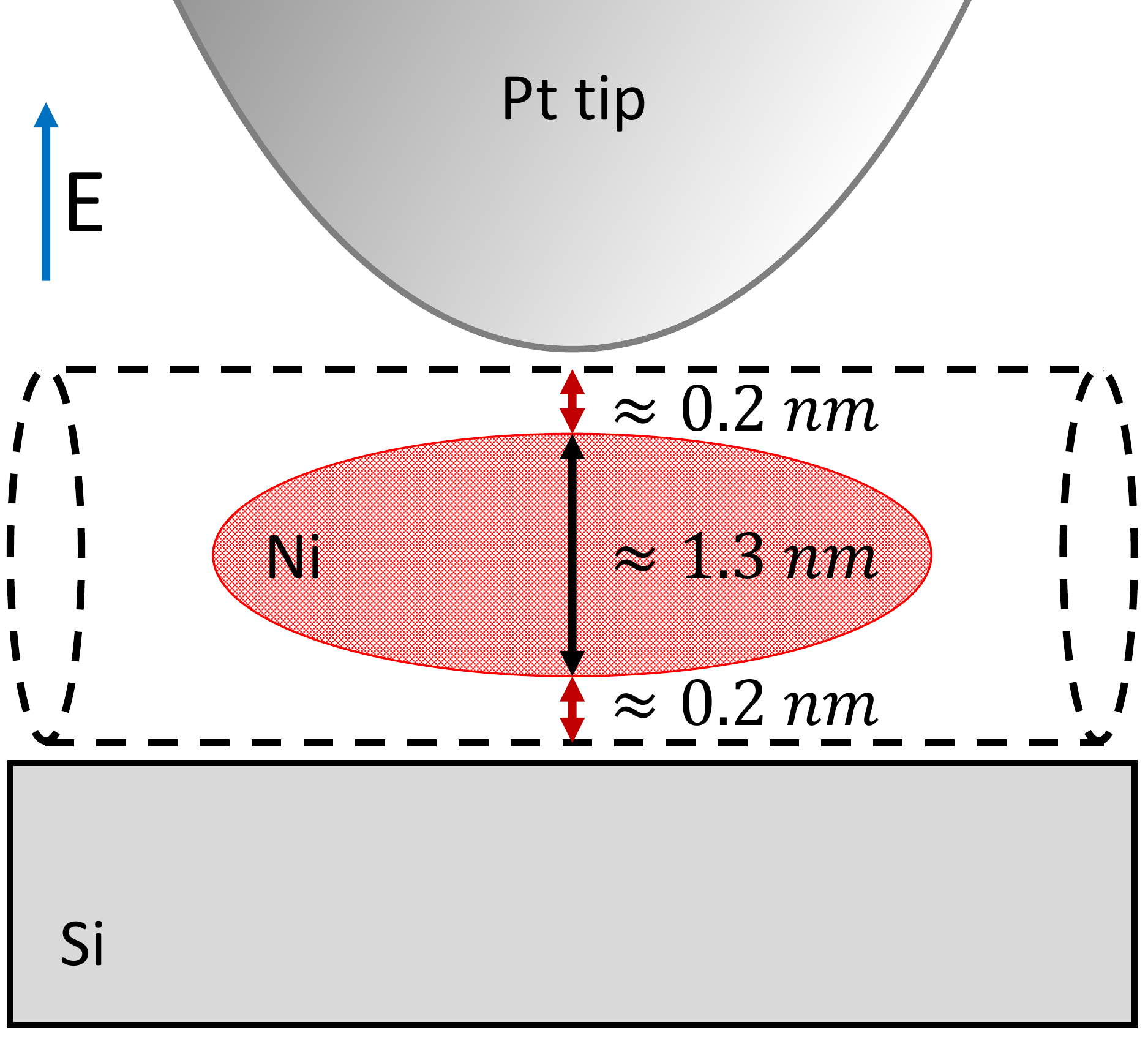}
\caption{Schematic representation of the probe-sample configuration used in our analytical model. The nickel cluster is modeled as a prolate spheroid. The nanotube wall (including the van der Waals distances) separates the tip and the substrate from the side of the nickel cluster. The average diameter of the e-Dips nanotubes is 1.7 nm thus the nickel cluster is considered to be 1.3 nm in diameter. }
\label{fig:efdmschema}
\end{center}
\end{figure}

In order to predict the near-field contrast we performed calculations based on the extended finite dipole model (EFDM)\cite{ocelic07,cvitko}. In the infrared region where the wavelength is much longer than the characteristic size of the tip-particle-substrate system, the scattering problem can be approximated as an electrostatic problem at each time step (Rayleigh scattering). The system is modeled as depicted in Fig. \ref{fig:efdmschema}. The tip is approximated as a prolate spheroid, the nanotube as an infinite long cylinder. The nickel clusters are also cylinder-like objects as previous studies \cite{shiozawa15} suggested. In order to fit them to the analytical model we replaced them with prolate spheroids because their polarizability is very close to that of a cylinder.

Neglecting the effect of the carbon nanotube (see above), the main part of the near-field interaction occurs between the probing tip, the nickel cluster and the substrate. The electric field is considered to be perpendicular to the surface of the substrate as the tip enhances the electric field component that is parallel to its axis of revolution. In such a model we can take into account the electric field response of the nickel cluster with several dipoles generated inside the nickel nanoparticle by the tip and the mirror charges of the substrate. Those dipoles are described by the local electric field and the polarizability of the object $(p=\alpha\cdot E_{loc})$. The polarizability of prolate spheroids, perpendicular to their major axis is given by the formula \cite{Venermo2005}:
\begin{equation}
\alpha_\perp=\frac{4}{3}R^2L\left(\frac{2R}{L}\right)^2\frac{\epsilon-1}{1+N_\perp (\epsilon-1)}
\label{eq:polariz}
\end{equation}
Here $\epsilon$ is the dielectric permittivity of the nanoparticle at the wavelength of interest, $R$ is the radius, and $L$ is the length of the elliptical nanoparticle, $N_\perp$ is the depolarization factor. The latter describes how much the internal field within the spheroid is attenuated by the polarization and it depends on the geometry of the object. This can be expressed as
\begin{equation}
N_\perp=\frac{1}{2}(1-N_\parallel)
\end{equation}
\begin{equation}
N_\parallel=\frac{1-e^2}{1e^3}\left(\ln \frac{1+e}{1-e}-2e\right)
\label{eq:depolfact}
\end{equation}
where $e=\sqrt{1-\left(\frac{R}{L/2}\right)}$ is the eccentricity of the spheroid. The diameter is chosen to be $D=$1.3 $nm$ to fit inside a nanotube with diameter 1.7 $nm$.
The dielectric permittivity of nickel was determined from the Drude model with a size-dependent term for the damping constant \cite{arboleda16}:
\begin{equation}
\epsilon(\omega,R)=1-\frac{\omega_p^2}{\omega^2+i\omega\gamma+i\omega C\frac{v_F}{R}}
\label{eq:drude}
\end{equation}
where $v_F$ is the Fermi velocity and $C$ is a factor that depends on the electron scattering process inside the particle. The Drude model parameters for nickel were taken from Ref. \cite{arboleda16}.

\begin{figure}[h!]
\begin{center}
\includegraphics[width=0.5\linewidth]{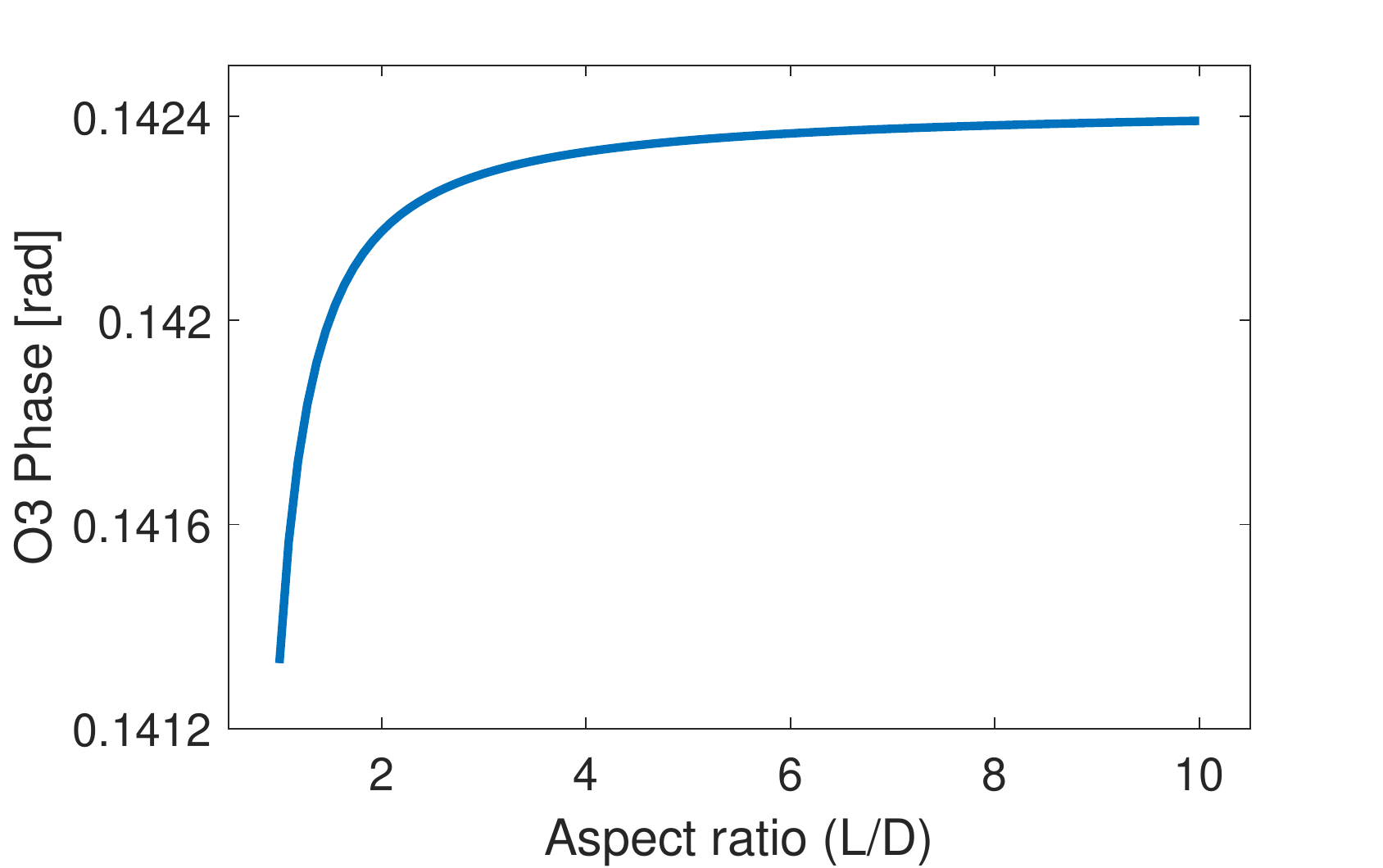}
\caption{Third harmonic (O3) near-field phase signal of a nickel nanoparticle  calculated from the EFDM model, as a function of aspect ratio $(L/D)$. All data were normalized to the signal from the silicon substrate.}
\label{fig:efdmres}
\end{center}
\end{figure}

These parameters were used in the EFDM model to calculate the 3rd harmonic demodulated near-field phase contrast of nickel nanoparticles compared to the silicon substrate versus the aspect ratio of the nanoparticle $(L/D)$. The result is presented in Fig. \ref{fig:efdmres}.

Fig. \ref{fig:efdmres} illustrates that the near-field phase contrast does not vary noticeably with the aspect ratio: the near-field contrast changes only by $\Delta\varphi_{O3}=0.001\, [rad]$ until it starts to saturate when the aspect ratio is around four. This amount of change cannot be detected with our instrument.

\section{Results and discussion}
\begin{figure}[h!]%
\centering
\subfloat{
\includegraphics*[width=0.4\linewidth]{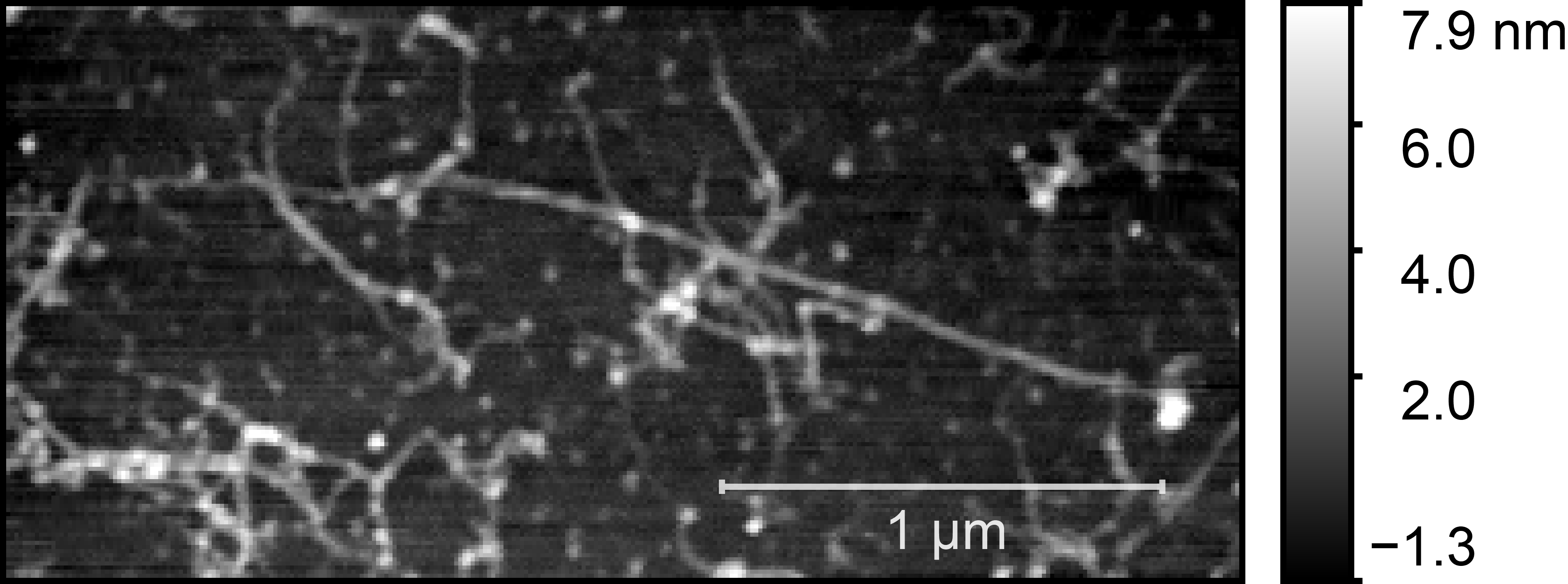}}
\subfloat{
\includegraphics*[width=0.4\linewidth]{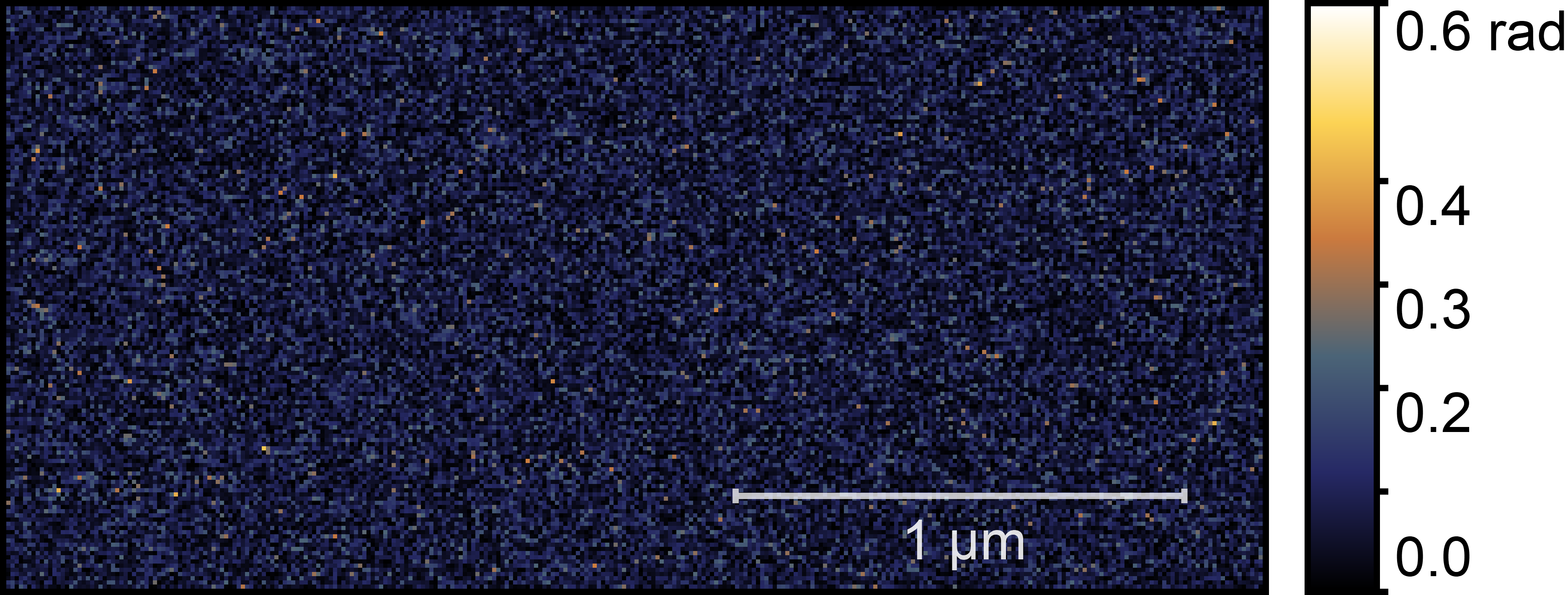}}
\caption{%
 AFM topography (left) and O3 near-field map (right) of a typical \Ni-filled carbon nanotube bundle before the annealing process. Optical images were taken with a $\nu$=980 cm$^{-1}$ illuminating laser.}
\label{fig:smallmapAP}
\end{figure}

Results on as-prepared \Ni -filled nanotubes transferred onto silicon substrate are shown in Fig. \ref{fig:smallmapAP}. We were looking for nanotube bundles with as small diameter as possible. We expect no phase signal from such bundles as \Ni\  molecular vibrations are too weak to provide an observable near-field signal. We found that nanotube bundles smaller than 10 $nm$ have no contrast on the near-field optical maps. The figure presents the AFM topography and the O3 near-field phase map of a nanotube bundle with diameter of $3nm$. Since the diameter of an individual e-Dips nanotube is $d\approx$ 1.7 $nm$,  bundles with $d=$3 $nm$ probably consist of three nanotubes. The lack of a near-field signal verifies our predictions.

\begin{figure}[h!]%
\centering
\subfloat{
\includegraphics*[width=0.25\linewidth]{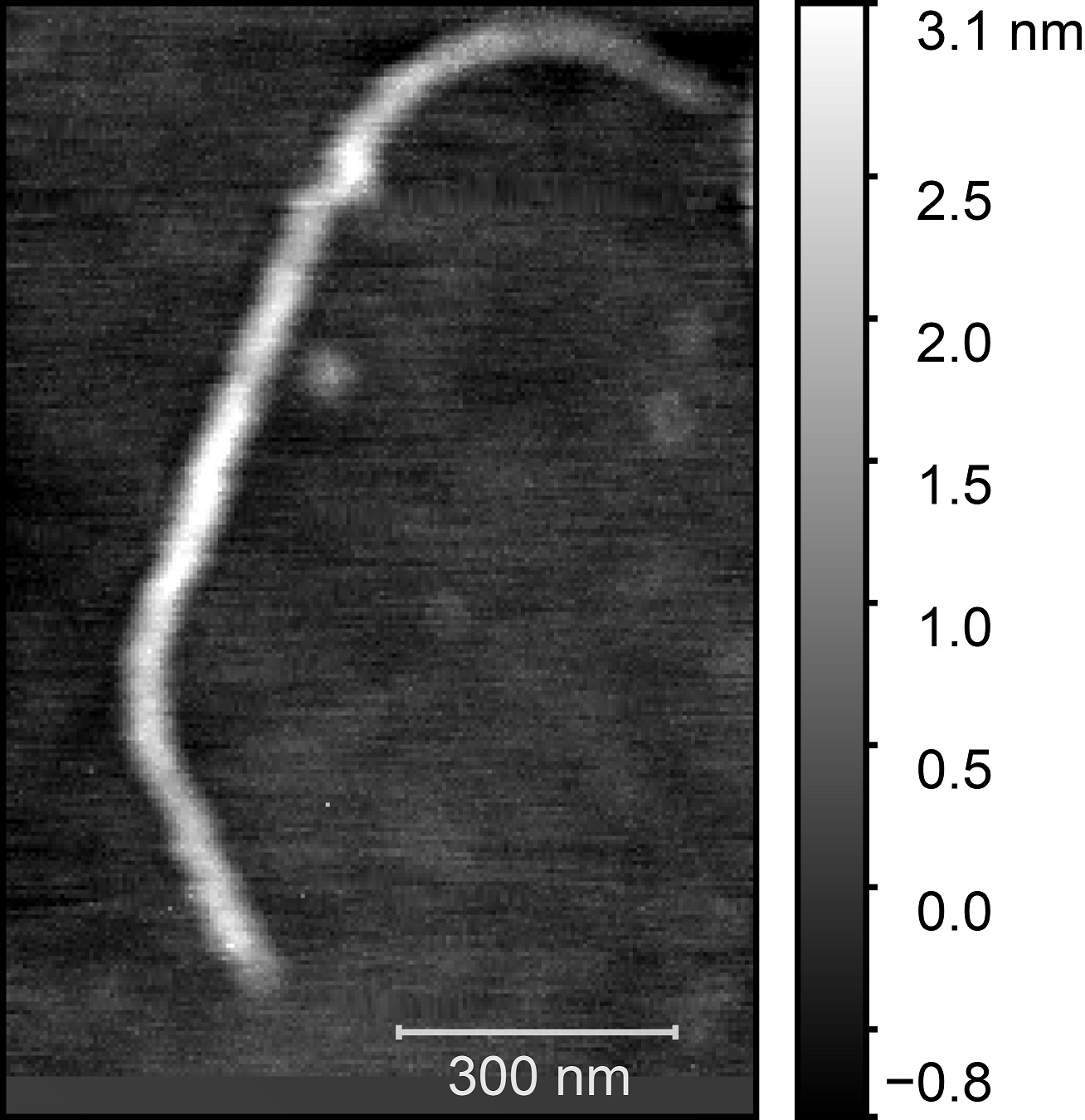}}
\subfloat{
\includegraphics*[width=0.25\linewidth]{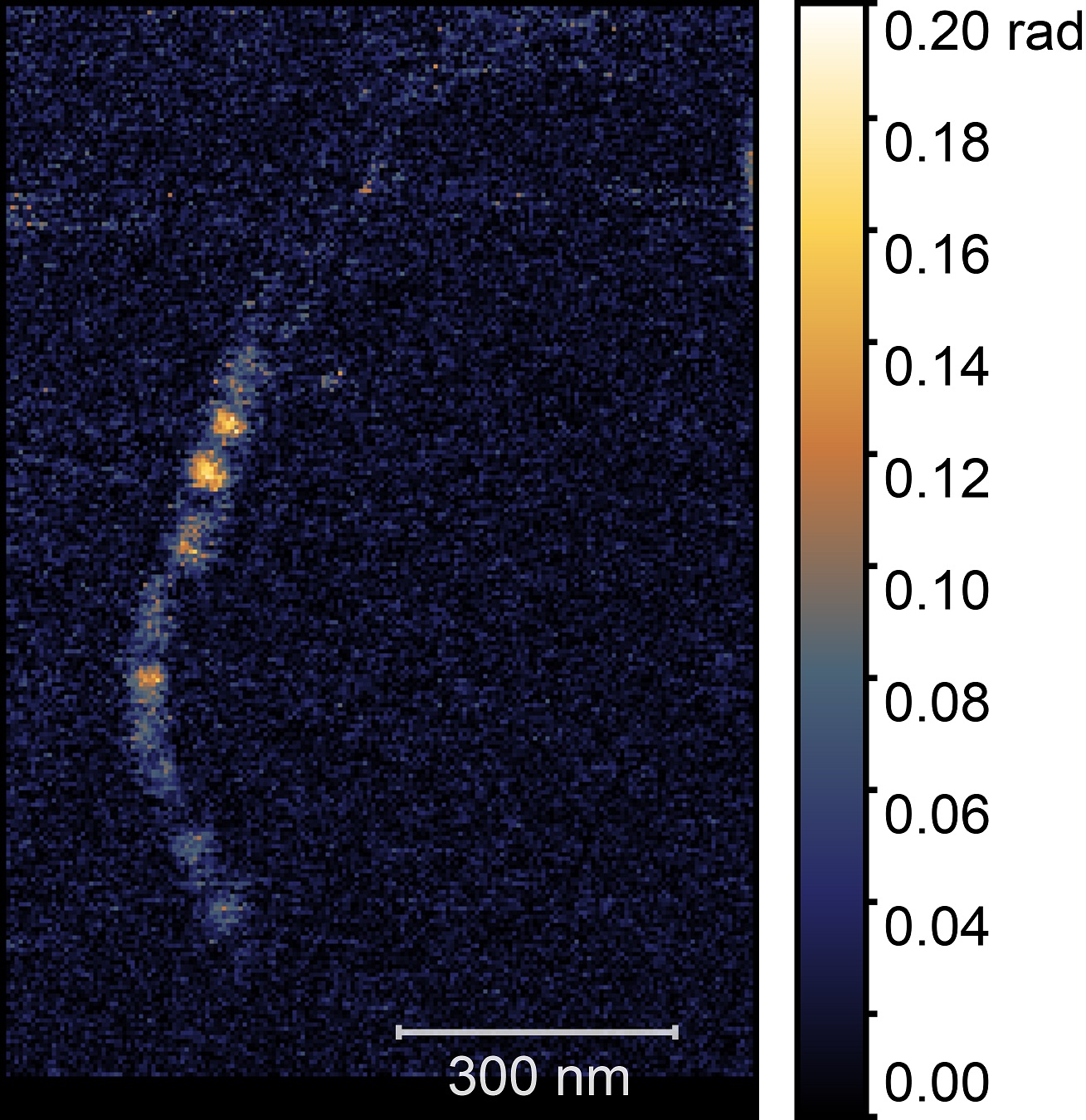}}
\caption{%
 AFM topography (left) and O3 near-field map (right) of a typical carbon nanotube bundle after the annealing process. Optical images were taken with a $\nu$=980 cm$^{-1}$ illuminating laser.  }
\label{fig:NiceTube}
\end{figure}

Next, we applied the annealing process ($700^\circ C, 2 h$) to this sample to create nickel clusters and repeated the optical characterization of the nanotubes. During the process the surface morphology of the sample changed substantially and it was impossible to find the same nanotube. However, we searched for nanotube bundles with identical diameter to the ones measured before. Fig \ref{fig:NiceTube} shows the AFM topography and the third harmonic phase signal of such a nanotube bundle.

\begin{figure}[h!]%
\centering
\subfloat{%
\includegraphics*[width=0.35\linewidth]{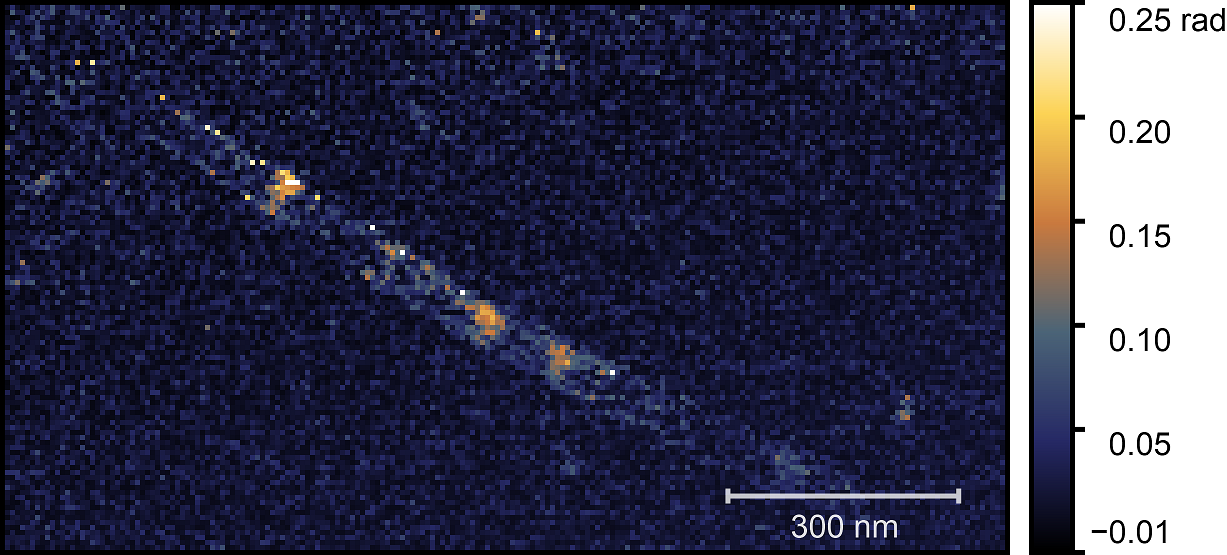}}
\subfloat{%
\includegraphics*[width=0.35\linewidth]{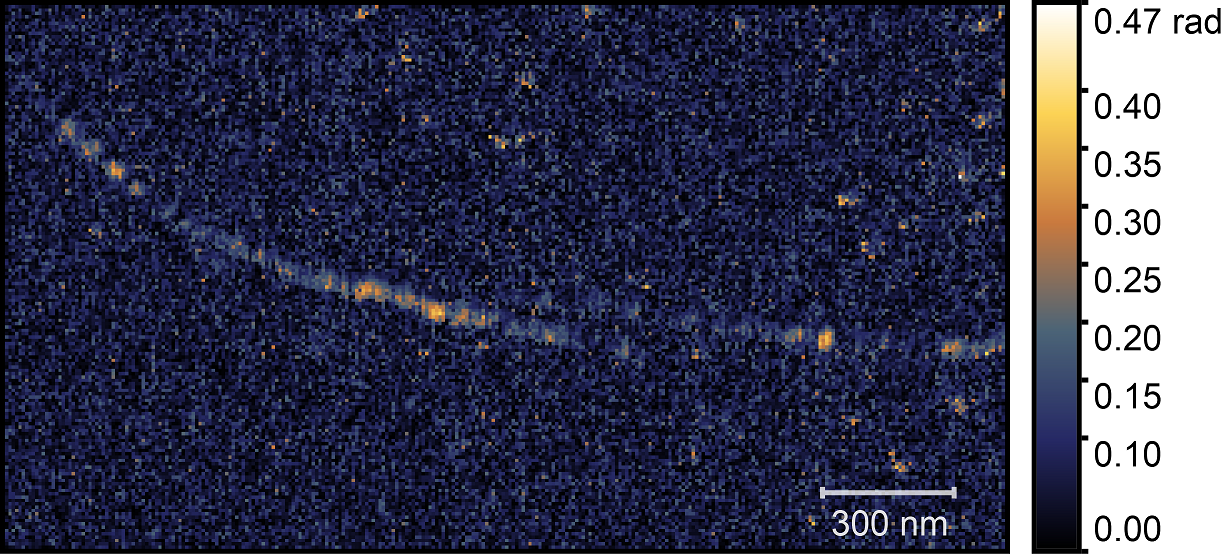}}
\caption{%
 O3 optical phase maps of two different carbon nanotube bundles. Images were taken with a $\nu$=980 cm$^{-1}$ illuminating laser. The presence of nickel clusters is very obvious as they cause high phase contrast.}
\label{fig:othermaps}
\end{figure}

As the optical image demonstrates, bright, high contrast spots appeared along the nanotubes. There are also places where no optical signal is found. That result corresponds to the transmission electron microscopy images which showed inhomogeneous spatial distribution of nickel clusters along the nanotube bundles. Our analytical model matches the measurements as well. Phase contrast values of the brightest spots are $\varphi_{O3}=0.139\pm 0.01 [rad]$. This agrees with a calculated value of $\varphi\approx 0.142 [rad]$. Other examples of phase maps can be seen in Fig. \ref{fig:othermaps}, where the location of the nickel clusters can be easily identified. We also found very few nanotubes with no contrast, which indicates that the filling of most tubes was sufficient to obtain nickel clusters. The contrast values vary within the sample; smaller values could correspond to smaller diameter clusters that do not fill the nanotube perfectly, larger contrast, in turn, could indicate multiple clusters measured together at the bottom of the tip.

\begin{figure}[h!]%
\centering
\includegraphics*[width=0.8\linewidth]{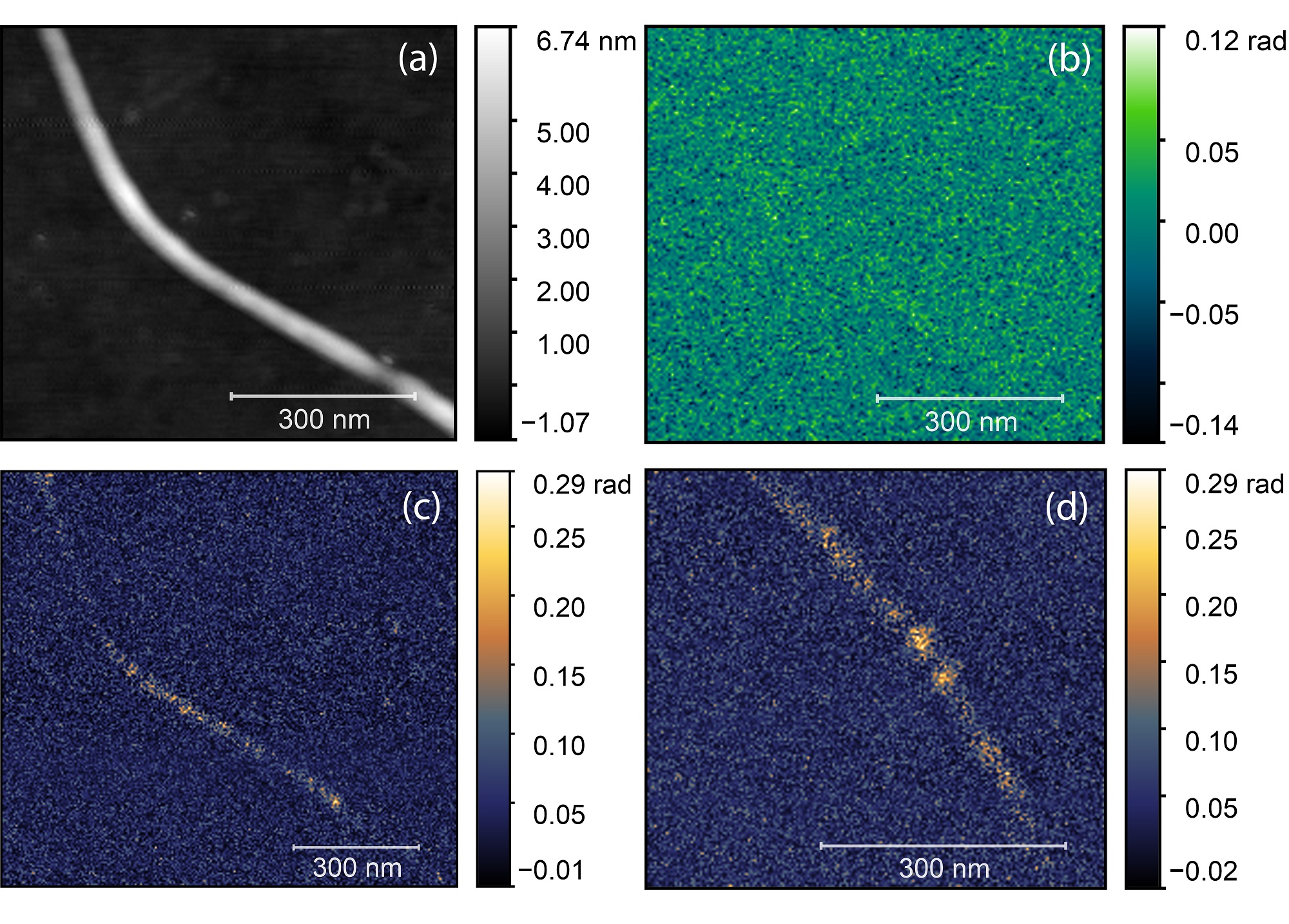}
\caption{%
 AFM topography (a), MFM phase (b), O3 near-field phase (c,d) of the same nanotube bundle. Figures clearly show contrast in optical images while magnetic microscopy is not sensitive to nickel clusters of this size.}
\label{fig:mfm}
\end{figure}

We also tried to localize the nickel clusters by magnetic force microscopy (MFM). This technique was already applied to investigate nanoparticles with magnetic behavior \cite{Garcia2002,Kinsella2007,Wang2009}. The excellent magnetic properties of these nickel clusters were already demonstrated \cite{shiozawa15}. We expected to obtain the signal of nickel clusters in the MFM phase images as magnetic dipoles give bright and dark spots at their opposite poles. Fig. \ref{fig:mfm} displays the AFM topography, the MFM and s-SNOM measurements on the same nanotube bundle. We repeated the MFM measurements with different tip lift height $(15,25,35,50,70,130nm)$ until the topographic related phase appeared. We did not find any sign of nickel clusters with MFM probing. If we assume that the aspect ratio of a nickel cluster is four (this was the saturation limit in the calculated near-field phase) and we treat it like a cylinder shaped cluster of face centered cubic (f.c.c.) structured nickel, we can roughly estimate the number of atoms measured in one spot. In order to do this, we used the lattice constant of f.c.c. nickel, $0.35\,nm$ \cite{davey25} and calculated how many cubes can fit in the above mentioned cylinder. This number was then multiplied by four because the unit cell of an f.c.c. crystal contains 4 atoms, giving the number of atoms to be around 644. We find that near-field optical probing is more sensitive and gives reliable information about the location of nickel clusters inside carbon nanotubes even in the case of such a small amount of material.

\section{Conclusions}
We observed nickel nanoclusters grown inside single walled carbon nanotubes via near-field microscopy based on their infrared free-carrier absorption. We found that these measurements are very sensitive to the presence of the metallic phase. We were able to detect optical signals from nickel atom clusters consisting of less than 700 nickel atoms. Our modified EFDM model gives phase contrasts close to the measured values and is found to reliably predict the optical signal of nanoparticles. We also detected nickel clusters in most of the nanotubes, consistent with electron microscopy results.


\section{Acknowledgement}
Research supported by the Hungarian National Research Fund (OTKA) No. SNN 118012. \'A.P. gratefully acknowledges support from the J\'anos Bolyai Fellowship of the Hungarian Academy of Sciences and from the National Research, Development and Innovation Office - NKFIH PD 121320 and NKFIH FK-125063. Transmission electron microscopy was supported by the VEKOP-2.3.3-15-2016-00002 project. H.S. acknowledges funding from the Austrian Science Fund (FWF) P30431-N36.

\bibliographystyle{unsrt}
\bibliography{G_Nemeth_ref-final}

\begin{thebibliography}{10}

\bibitem{sloan98}
J.~Sloan, J.~Hammer, M.~Zwiefka-Sibley, and M.L.H. Green.
\newblock The opening and filling of single walled carbon nanotubes {(SWTs)}.
\newblock {\em Chem. Commun.}, 1998:347--348, 1998.

\bibitem{soldano15}
Caterina Soldano.
\newblock Hybrid metal-based carbon nanotubes: {Novel} platform for
  multifunctional applications.
\newblock {\em Prog. Mater. Sci.}, 69:183--212, 2015.

\bibitem{guan08}
Lunhui Guan, Kazu Suenaga, Shingo Okubo, Toshiya Okazaki, and Sumio Iijima.
\newblock Metallic wires of lanthanum atoms inside carbon nanotubes.
\newblock {\em J. Am. Chem. Soc.}, 130:2162--2163, 2008.

\bibitem{lutz10}
M.U. Lutz, U.~Weissker, F.~Wolny, C.~{M\"uller}, M.~{L\"offler}, T.~{M\"uhl},
  A.~Leonhardt, B.~{B\"uchner}, and R.~Klingeler.
\newblock Magnetic properties of {$\alpha$-Fe} and {FE$_3$C} nanowires.
\newblock {\em J. Phys.: Conf. Ser.}, 200:072062--1--4, 2010.

\bibitem{jankovic06}
Lubo{\v s} Jankovi{\v c}, Dimitrios Gournis, Pantelis~N. Trikalitis, Imad
  Arfaoui, Tristan Cren, Petra Rudolf, {Marie-H\'elene} Sage, {Thomas T.M.
  Palstra}, Bart Kooi, Jeff {De Hosson}, Michael~A. Karakassides, Konstantinos
  Dimos, Aliki Moukarika, and Thomas Bakas.
\newblock Carbon nanotubes encapsulating superconducting single-crystalline tin
  nanowires.
\newblock {\em Nano Lett.}, 6:1131--1135, 2006.

\bibitem{tombros08}
Nikolaos Tombros, Luuk Buit, Imad Arfaoui, Theodoros Tsoufis, Dimitrios
  Gournis, Pantelis~N. Trikalitis, Sense~Jan {van der Molen}, Petra Rudolf, and
  Bart~J. {van Wees}.
\newblock Charge transport in a single superconducting tin nanowire
  encapsulated in a multiwalled carbon nanotube.
\newblock {\em Nano Lett.}, 8:3060--3064, 2008.

\bibitem{li05b}
L.-J. Li, A.~N. Khlobystov, J.~G. Wiltshire, G.~A.~D. Briggs, and R.~J.
  Nicholas.
\newblock Diameter-selective encapsulation of metallocenes in single-walled
  carbon nanotubes.
\newblock {\em Nat. Mater.}, 4:481--485, 2005.

\bibitem{li06}
Y.~F. Li, R.~Hatakeyama, T.~Kaneko, T.~Izumida, T.~Okada, and T.~Kato.
\newblock Synthesis and electronic properties of ferrocene-filled double-walled
  carbon nanotubes.
\newblock {\em Nanotechnology}, 17:4143--4147, 2006.

\bibitem{shiozawa08}
Hidetsugu Shiozawa, Thomas Pichler, Alexander Gr{\"u}neis, Rudolf Pfeiffer,
  Hans Kuzmany, Zheng Liu, Kazu Suenaga, and Hiromichi Kataura.
\newblock A catalytic reaction inside a single-walled carbon nanotube.
\newblock {\em Adv.Mater.}, 20:1443--1449, 2008.

\bibitem{shiozawa15}
Hidetsugu Shiozawa, Antonio Briones-Leon, Oleg Domanov, Georg Zechner, Yuta
  Sato, Kazu Suenaga, Takeshi Saito, Michael Eisterer, Eugen Weschke, Wolfgang
  Lang, Herwig Peterlik, and Thomas Pichler.
\newblock Nickel clusters embedded in carbon nanotubes as high performance
  magnets.
\newblock {\em Sci. Rep.}, 5:15033--1--8, 2015.

\bibitem{wu04}
Zhuangchun Wu, Zhihong Chen, Xu~Du, Jonathan~M. Logan, Jennifer Sippel, Maria
  Nikolou, Katalin Kamaras, John~R. Reynolds, David~B. Tanner, Arthur~F.
  Hebard, and Andrew~G. Rinzler.
\newblock Transparent, conductive carbon nanotube films.
\newblock {\em Science}, 305:1273--1276, 2004.

\bibitem{Bharadwaj2009}
Palash Bharadwaj, Bradley Deutsch, and Lukas Novotny.
\newblock Optical antennas.
\newblock {\em Adv. Opt. Photon.}, 1:438--483, 2009.

\bibitem{Hillenbrand2002}
R.~Hillenbrand and F.~Keilmann.
\newblock Material-specific mapping of metal/semiconductor/dielectric
  nanosystems at 10 nm resolution by backscattering near-field optical
  microscopy.
\newblock {\em Applied Physics Letters}, 80:25--27, 2002.

\bibitem{Hillenbrand2000}
R.~Hillenbrand and F.~Keilmann.
\newblock Complex optical constants on a subwavelength scale.
\newblock {\em Phys. Rev. Lett.}, 85:3029--3032, 2000.

\bibitem{Stiegler2011}
J.~M. Stiegler, Y.~Abate, A.~Cvitkovicand, Y.~E. Romanyuk, A.~J. Huber, S.~R.
  Leone, and R.~Hillenbrand.
\newblock {Nanoscale Infrared Absorption Spectroscopy of Individual
  Nanoparticles Enabled by Scattering-Type Near-Field Microscopy}.
\newblock {\em ACS Nano}, 5:9494--6499, 2011.

\bibitem{Hillenbrand2006}
R.~Hillenbrand and F.~Keilmann.
\newblock Pseudoheterodyne detection for background-free near-field
  spectroscopy.
\newblock {\em Appl. Phys. Lett.}, 89:101124--1--3, 2006.

\bibitem{nemeth16}
G.~N\'emeth, D.~Datz, H.~M. T\'oh\'ati, {\'A}.~Pekker, and K.~Kamar\'as.
\newblock {Scattering near-field optical microscopy on metallic and
  semiconducting carbon nanotube bundles in the infrared}.
\newblock {\em physica status solidi (b)}, 253:2413--2416, 2016.

\bibitem{nemeth17}
G.~N\'emeth, {\'A}.~Pekker, D.~Datz, H.M. T\'oh\'ati, K.~Otsuka, T.~Inoue,
  S.~Maruyama, and K.~Kamar\'as.
\newblock {Nanoscale characterization of individual horizontally aligned
  single-walled carbon nanotubes}.
\newblock {\em physica status solidi (b)}, 254:1700433--1--4, 2017.

\bibitem{ocelic07}
N.~Ocelic.
\newblock {\em Quantitative Near-field Phonon-polariton Spectroscopy}.
\newblock PhD thesis, Technical University Munich, (2007).

\bibitem{cvitko}
A.~Cvitkovic.
\newblock {\em Substrate-Enhanced Scattering-Type Scanning Near-Field Infrared
  Microscopy of Nanoparticles}.
\newblock PhD thesis, Technische {Universit\"at} {M\"unchen}, (2009).

\bibitem{Venermo2005}
J.~Venermo and A.~Sihvola.
\newblock {Dielectric polarizability of circular cylinder}.
\newblock {\em Journal of Electrostatics}, 63:101--117, 2005.

\bibitem{arboleda16}
David~Muneton Arboleda, Jesica M~J Santillan, Luis J~Mendoza Herrera, Diego
  Muraca, Daniel~C Schinca, and Lucia~B Scaffardi.
\newblock Size-dependent complex dielectric function of {Ni, Mo, W, Pb, Zn and
  Na} nanoparticles. {Application} and sizing.
\newblock {\em J. Phys. D}, 49:075302, 2016.

\bibitem{Garcia2002}
J.M Garcia, A~Thiaville, and J~Miltat.
\newblock Mfm imaging of nanowires and elongated patterned elements.
\newblock {\em Journal of Magnetism and Magnetic Materials}, 249:163 -- 169,
  2002.

\bibitem{Kinsella2007}
Joseph~M. Kinsella and Albena Ivanisevic.
\newblock Dna-templated magnetic nanowires with different compositions:?
  fabrication and analysis.
\newblock {\em Langmuir}, 23:3886--3890, 2007.
\newblock PMID: 17316030.

\bibitem{Wang2009}
T~Wang, Y~Wang, Y~Fu, T~Hasegawa, F~S Li, H~Saito, and S~Ishio.
\newblock A magnetic force microscopy study of the magnetic reversal of a
  single fe nanowire.
\newblock {\em Nanotechnology}, 20:105707, 2009.

\bibitem{davey25}
Wheeler~P. Davey.
\newblock Precision measurements of the lattice constants of twelve common
  metals.
\newblock {\em Phys. Rev.}, 25:753--761, 1925.

\end{thebibliography}

	
\end{document}